\documentclass[aps,prl,twocolumn,tightenlines,superscriptaddress,showpacs,fleqn]{revtex4}
\usepackage{esvect}
\usepackage{amssymb}
\usepackage{graphicx}
\usepackage{amsmath}
\usepackage{amsbsy}
\usepackage{amsthm}
\usepackage[ansinew]{inputenc} 
\usepackage{bbm}
\usepackage{bm}
\usepackage{epsfig}
\usepackage{lscape}
\usepackage{bbold}
\usepackage{hhline}

\newcommand{\beq}{\begin{eqnarray}}
\newcommand{\eeq}{\end{eqnarray}}

\usepackage[varg]{txfonts}

\begin{document}

\title{
Weak Interaction Processes: Which Quantum Information is revealed?}

\author{B. C. Hiesmayr}
\affiliation{University of Vienna, Faculty of Physics, Boltzmanngasse 5, 1090 Vienna, Austria}


\begin{abstract}
We analyze the achievable limits of the quantum information processing of the weak interaction revealed by hyperons with spin. We find that the weak decay process corresponds to an interferometric device with a fixed visibility and fixed phase difference for each hyperon. Nature chooses rather low visibilities expressing a preference to parity conserving or violating processes (except for the decay $\Sigma^+\longrightarrow p \pi^0$). The decay process can be considered as an open quantum channel that carries the information of the hyperon spin to the angular distribution of the momentum of the daughter particles. We find a simple geometrical information theoretic interpretation of this process: two quantization axes are chosen spontaneously with probabilities $\frac{1\pm\alpha}{2}$ where $\alpha$ is proportional to the visibility times the real part of the phase shift. Differently stated the weak interaction process corresponds to spin measurements with an imperfect Stern-Gerlach apparatus. Equipped with this information theoretic insight we show how entanglement can be measured in these systems and why Bell's nonlocality (in contradiction to common misconception in literature) cannot be revealed in hyperon decays. We study also under which circumstances contextuality can be revealed.
\end{abstract}


\maketitle

Weak interactions are one out of the four fundamental interactions that we think that rules our universe. The weak interaction is the only interaction that breaks the parity symmetry and the combined charge-conjugation--parity ($\mathcal{C}P$) symmetry. Recently, it has been shown that for K-mesons~\cite{Hiesmayr:2012}, which are spinless particles decaying via the weak interaction, Bell's theorem can be related to the violation of the $\mathcal{C}P$ symmetry. Herewith one of the most counterintuitive property of quantum mechanics relates itself to the unsolved problem of the matter-antimatter asymmetry in our universe. This paper focus on particles having half-integer spin and decaying via the weak interaction, i.e. hyperons that are baryons containing in addition to up or down quarks also one or more strange quarks. Hyperon decays violate the parity ($P$) symmetry. The Standard Model of elementary particles predicts also tiny contribution of  $\mathcal{C}P$ violating processes, however, no violation of the $\mathcal{C}P$ symmetry has been up to now experimentally found.

We develop an information theoretic description of the weak decay process of hyperons which also puts the new expected data (e.g. by the PANDA experiment~\cite{Lutz:2009ff} at FAIR in Germany and BES-III~\cite{Asner:2008nq} at BEPC II in China) into a unified picture, connects it to quantum information theory and clears some misunderstandings existing in the literature. In particular, we find that the $P$ violating and non-violating amplitudes correspond to the two arms of an asymmetric interferometer allowing for a quantified discussion of Bohr's complementarity relation. We can further show that the decay process is a noisy quantum channel or differently stated, the spin of the decaying particle is only imperfectly measured by the decay process. Equipped with this quantum information theoretic insight we proceed to the two-particle case where experimental data suggests that there is entanglement in the system. We introduce an optimal observable that witnesses the entanglement in the spin degrees of freedom. Last but not least we discuss whether a test versus local realistic theories via Bell inequalities is possible and whether the contextuality property of the quantum theory can be revealed in hyperon-antihyperon systems.

\textbf{Information theoretic content of an interfering and decaying system:}
Let us start by assuming that the initial state of a decaying hyperon is in a separable state between momentum and spin degrees of freedoms $
\rho_{\textrm{hyperon}}=\rho_{mom}\otimes\rho_{spin}$.
Typically one is interested in the angular distribution of the decay products $(\theta,\phi)$ depending on the polarization state of the hyperon particle, i.e.
a projection onto the above state
\beq
I(\theta,\phi)&=&
Tr_{spin} Tr_{mom}\; |\Psi\rangle\langle\Psi|\; \rho_{mom}\otimes\rho_{spin}\nonumber\\
&=& Tr_{spin}\; T\;\rho_{spin}\;T^\dagger\;\nonumber
\eeq
where $T$ is usually dubbed decay or transition matrix. Often the final state can have different angular momenta, therefore the projector can be a superposition of the respective different angular momentum states which we denote by $a,b,c,\dots$. Assuming two different ones we obtain
\beq\label{generalresult}
\lefteqn{I(\theta,\phi)
=Tr_{spin}\;  (T_a+T_b)\;\rho_{spin}\;(T_a+T_b)^\dagger}\\
&=&(\|T_{a}\|^2+\|T_{b}\|^2)\cdot\nonumber\\
\lefteqn{\quad\biggl\lbrace 1+ 2 \frac{\|T_{a}\|\cdot \|T_{b}\|}{\|T_{a}\|^2+ \|T_{b}\|^2} Re\{Tr_{spin}\; \hat{T}_a\;\rho_{spin}\;\hat{T}_b^\dagger\}\biggr\rbrace\;,}\nonumber
\eeq
where we considered for the last equation the normalized quantities $\hat{T}_{a/b}=T_{a/b}\|T_{a,b}\|^{-1}$. The coefficient in front of the interference term can be considered as the visibility $\mathcal{V}$, the interference contrast of the two interfering amplitudes. A predictability $\mathcal{P}$ can be derived by computing the difference between the probability that a particle decays via the process $a$ and $b$
\beq\label{predictabilityvisibility}
\mathcal{P}=\left|\frac{\|T_{a}\|- \|T_{b}\|}{\|T_{a}\|^2+ \|T_{b}\|^2}\right|\;\textrm{and}\;\; \mathcal{V}=
2 \frac{\|T_{a}\|\cdot \|T_{b}\|}{\|T_{a}\|^2+ \|T_{b}\|^2}\;,
\eeq
and, obviously, we have
\beq
\mathcal{V}^2+\mathcal{P}^2\;=\;1\;,
\eeq
which is a quantitative rephrase of Bohr's complementarity relation or the closely related concept of duality in interferometric
devices. Predictability and visibility is a pair of complementary properties,
so the better we know one of them, the less we can determine the other one. One may relate predictability to the ``\textit{particle-like property}'', i.e. a ``\textit{which way}''--information, and visibility to the ``\textit{wave-like property}'', i.e. the interference contrast or sharpness. The first step in bringing the qualitative statement ``\textit{the observation of an interference pattern and the acquisition of which-way information are mutually exclusive}''
into a quantitative one was taken by Greenberger and Yasin~\cite{GreenbergYasin} and refined by Englert~\cite{Englert}. The authors of Ref.~\cite{Complementarity} investigated physical situations for which the complementary expressions depend only linearly on some variable $y$. This included interference patterns of various types of double slit experiments ($y$ is linked to position), but also oscillations due to particle mixing ($y$ is linked to time), e.g. by the neutral K-meson system, and also Mott scattering experiments of identical
particles or nuclei ($y$ is linked to a scattering angle). For the K-meson system the effect of the $\mathcal{C}P$ symmetry violation was investigated~\cite{ComplementarityCP} showing that it shifts obtainable information about our reality to different aspects,
without violating the complementarity principle, i.e. from predictability $\mathcal{P}$ to visibility $\mathcal{V}$ and vice versa.  All these two-state systems belonging to distinct fields of physics can then be treated via the generalized complementarity relation in a unified way. In Ref.~\cite{ComplentarityUnstable} the authors investigated how the un-stability due to a decay within the interferometer reduces the visibility.

\begin{figure}
\center{(a)\includegraphics[width=0.4\textwidth,keepaspectratio=true]{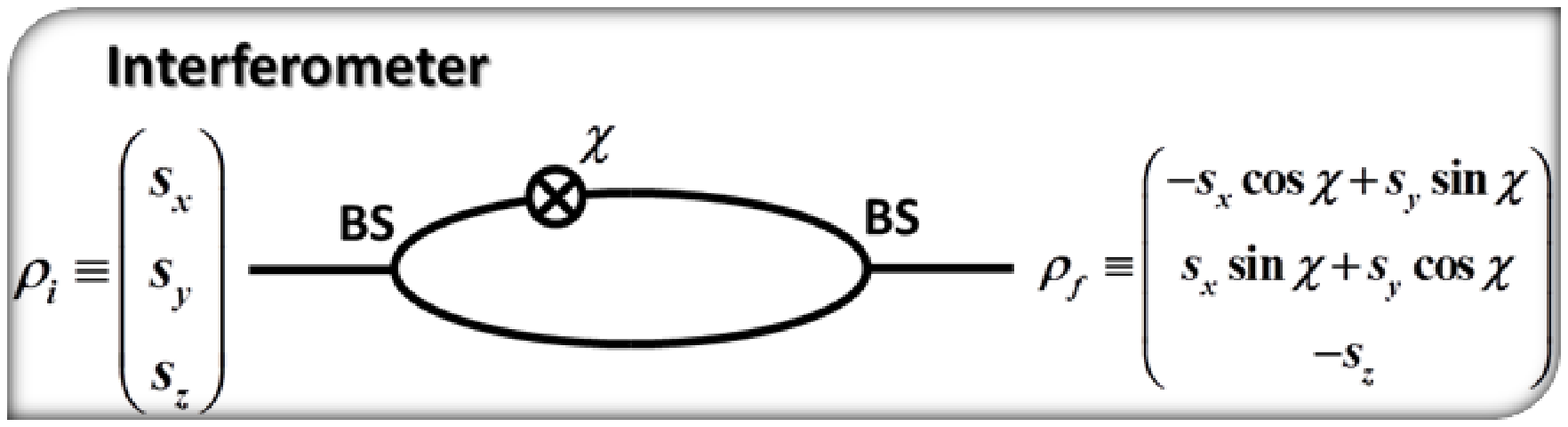}
(b)\includegraphics[width=0.4\textwidth,keepaspectratio=true]{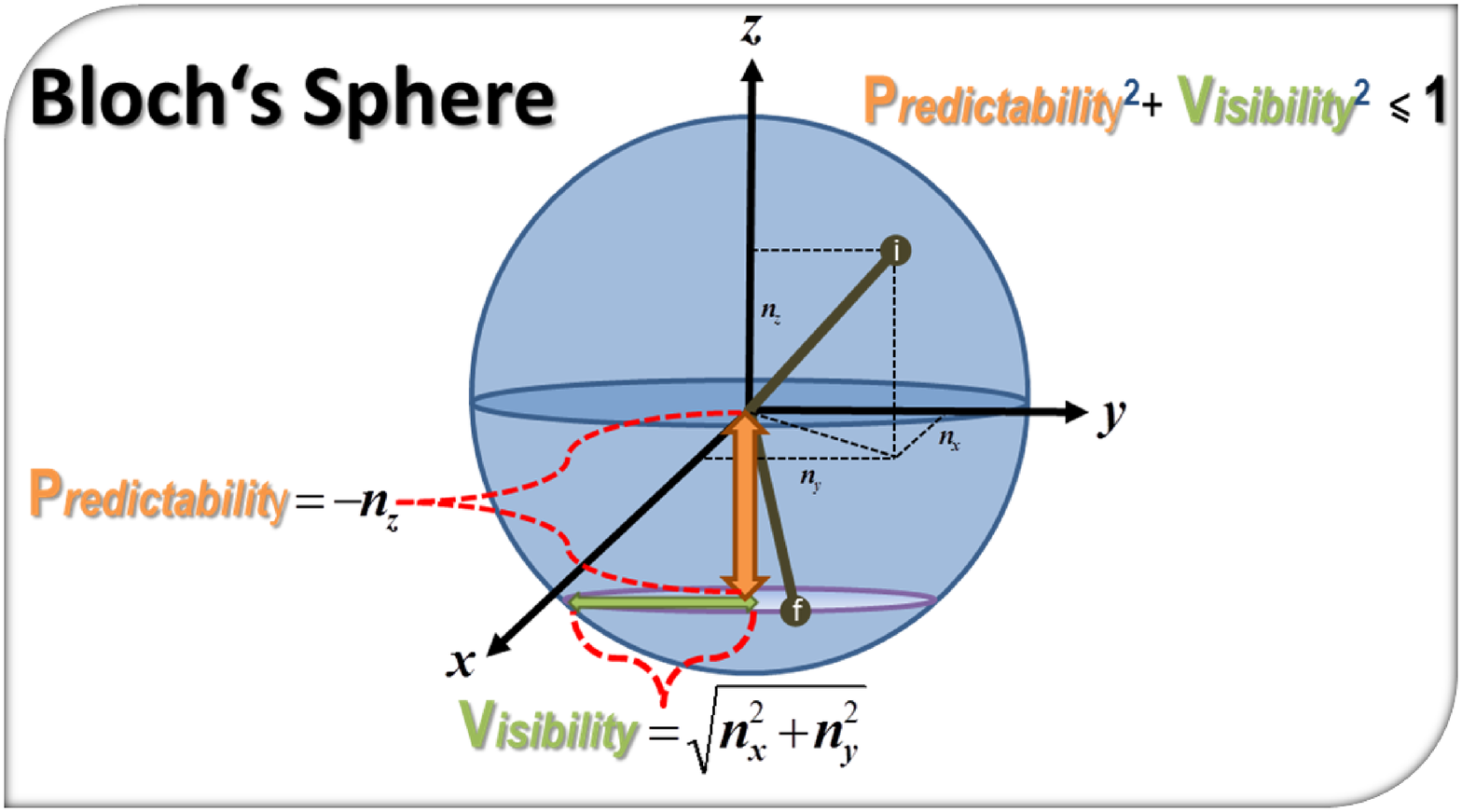}\caption{(Color online) (a) depicts an interferometer for a spin-$\frac{1}{2}$ particle with spin oriented along $\protect\vv{s}$ propagating through two beam-splitters (BS) and picking up a relative phase $\chi$. In (b) the Bloch's sphere is depicted and shows how a given interferometric setup changes the initial spin. Depending on the initial spin the interferometer reveals ``\textit{wave-particle}'' property, quantified by the visibility, or ``\textit{which-way}'' information, quantified by the predictability. Both quantities add up to one for pure spin states.
}\label{Blochsphere}}
\end{figure}

\begin{figure}
\center{
(a)\includegraphics[width=0.2\textwidth,keepaspectratio=true]{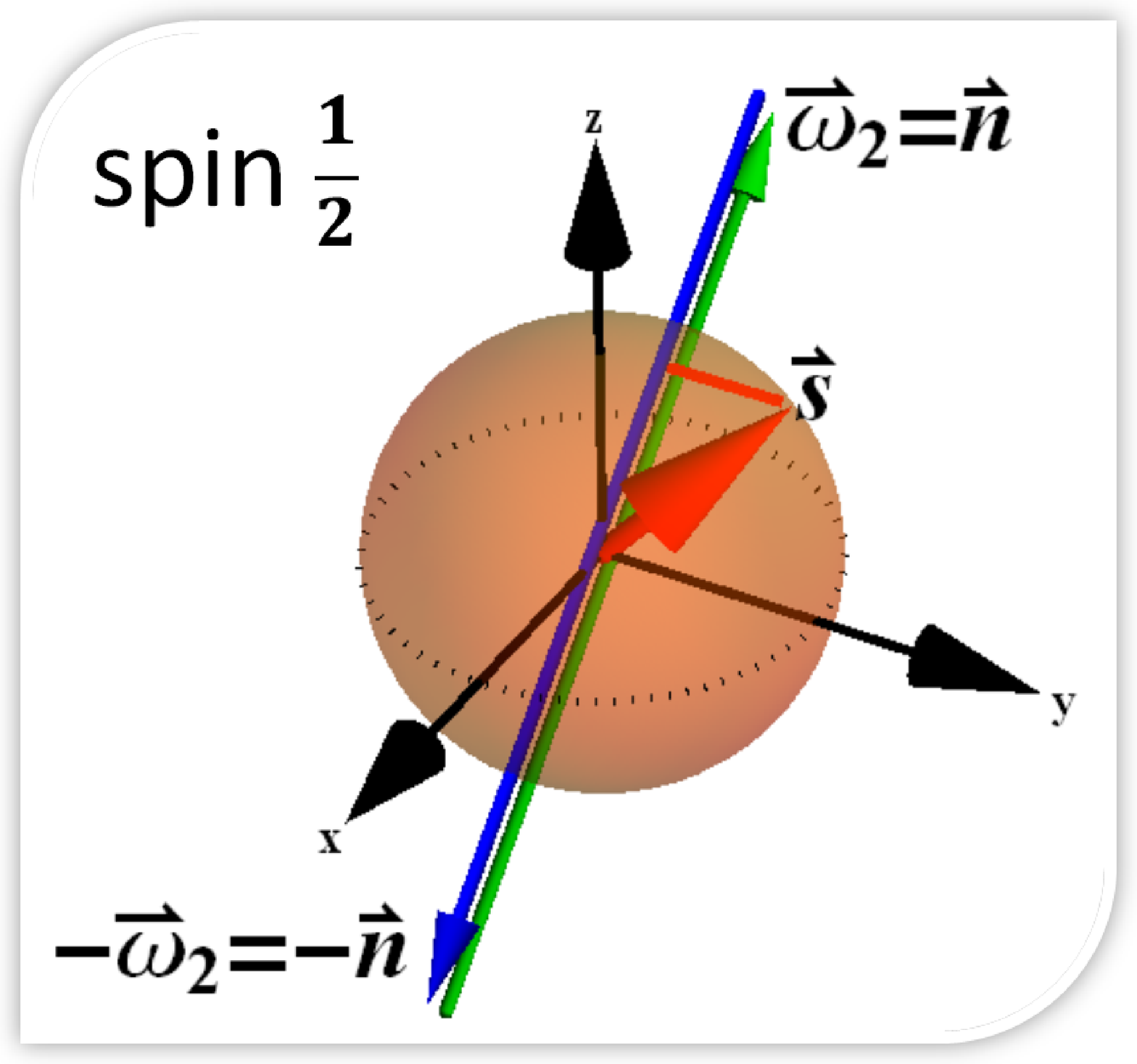}
(b)\includegraphics[width=0.2\textwidth,keepaspectratio=true]{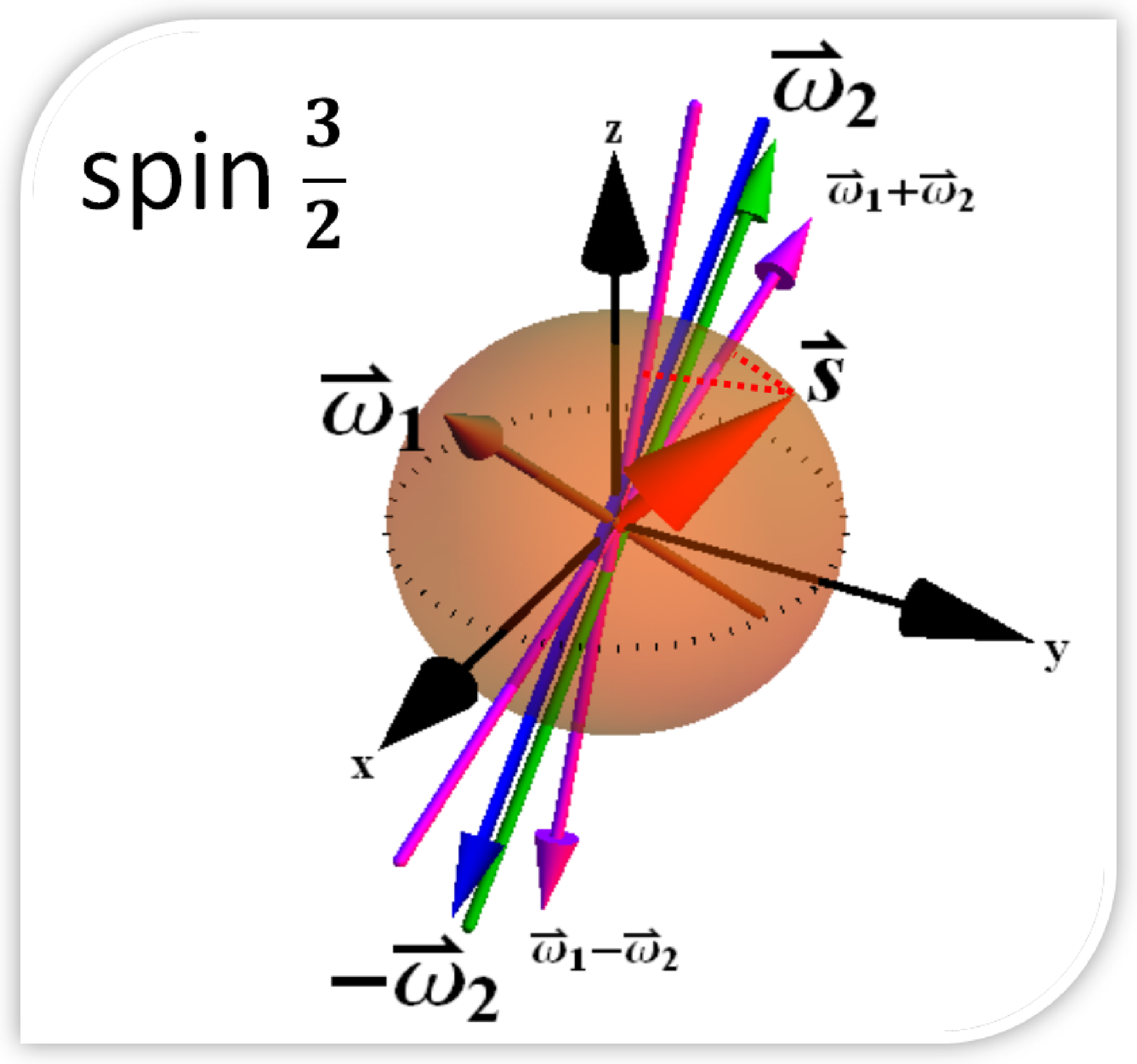}
\caption{(Color online) (a) shows the three dimensional Bloch's sphere for spin-$\frac{1}{2}$ particles and (b) the eight dimensional Bloch's sphere for spin-$\frac{3}{2}$ particles. Weak interaction depicts two quantization directions $\protect\vv{\omega}_1\pm\protect\vv{\omega}_2$ along which the initial spin of the hyperon is projected. For spin-$\frac{1}{2}$ particles $\protect\vv{\omega}_1=0$ and with probability $\omega_\pm$ the hyperon spin is projected onto $\pm\protect\vv{n}$ where $\protect\vv{n}$ corresponds to the momentum direction of the daughter particle. In case of spin-$\frac{3}{2}$ particles the real and imaginary parts of the transition amplitude differ resulting in a contribution to the quantization directions that is independent of difference of the two probabilities $\omega_\pm$. Hence the hyperon spin is projected onto two slightly tilted quantization directions $\protect\vv{\omega}_1\pm\protect\vv{\omega}_2$.
}\label{weakdecay}}
\end{figure}

\textbf{A simple open quantum formalism for decaying hyperons:} Quite generally, we can rewrite the dynamics in the spin space by switching to the formalism of open quantum systems  that allows a fast computation and transparent interpretation. In particular, we can always find two hermitian Kraus operators $K_{\pm}$ being the sum of the real and imaginary part of the decay matrix such that the following equation holds under the trace operation
\beq
I(\theta,\phi)&=&Tr_{spin}\;  (T_a+T_b)\;\rho_{spin}\;(T_a+T_b)^\dagger\nonumber\\
&\equiv & Tr_{spin}(K_+\rho_{spin} K_++K_-\rho_{spin} K_-)\nonumber
\eeq
where the Kraus operators have the conceptually simple form ($\omega_\pm>0$)
\beq
K_\pm&=\sqrt{\omega_{\pm}}\; |\vv{\omega}_1\pm\vv{\omega}_2\rangle\langle \vv{\omega}_1\pm\vv{\omega}_2|\,:=\, \sqrt{\omega_{\pm}}\; \Pi_{\vv{\omega}_1\pm\vv{\omega}_2}\nonumber
\eeq
with $\omega_++\omega_-=1$. The two Blochvectors have to be orthogonal, $\vv{\omega}_{1}\cdot\vv{\omega}_{2}=0$, since the transition is completely positive and are chosen such that they have maximal length $|\vv{\omega}_1\pm\vv{\omega}_2|^2=s (2s+1)$ ($s$\dots spin number). A Blochvector expansion of a density matrix is generally given by $\rho=\frac{1}{d}\{\mathbbm{1}_d+\vv{b}\cdot\vv{\Gamma}\}$ where $d$ is the dimension of the system~\cite{KrammerBertlmann}. Since we are dealing with spin-degrees of freedom we have $d=2s+1$ and we can choose as a set of orthonormal basis the generalized Hermitian and traceless Gell-Mann matrices  $\vv{\Gamma}$  (for $s=\frac{1}{2}$ they correspond to the Pauli matrices). Given this structure we can reinterpret the weak decay process as an incomplete spin measurement of the decaying particle
\beq\label{ErgInt}
I(\theta,\phi)&=&\omega_+\; Tr (\Pi_{\vv{\omega}_1+\vv{\omega}_2}\,\rho_{spin})+\omega_-\; Tr (\Pi_{\vv{\omega}_1-\vv{\omega}_2}\,\rho_{spin})\nonumber\\
&=&\frac{1}{(2s+1)}\{1+(\vv{\omega_1}+(\omega_+-\omega_-)\,\vv{\omega_2})\cdot\vv{s}\}
\eeq
where $\vv{s}$ is the Bloch vector representation of $\rho_{spin}$, i.e. $\vv{s}=Tr(\vv{\Gamma} \rho_{spin})$.
With probability $\omega_+$ the spin state of the hyperon is projected onto direction $\vv{\omega}_1+\vv{\omega}_2$ or with the remaining probability $\omega_-$ the initial spin state is measured along the direction $\vv{\omega}_1-\vv{\omega}_2$. Thus the weak process can be associated to a spin measurement with an imperfect Stern-Gerlach apparatus (switching with probability $\omega_{\pm}$ the magnetic field) which is geometrically depicted in Fig.~\ref{weakdecay}. The imperfection has two causes: Firstly, the difference $(\omega_+-\omega_-)$ equals an asymmetry (denoted in the following by $\alpha$) which corresponds to the characteristics of the interferometer. Secondly, the two directions $\vv{\omega}_1\pm\vv{\omega}_{2}$ are characteristical for the weak decay. Explicit examples are given below.

Let us remark here that for a spin-$\frac{1}{2}$ particle the decay dynamics simplify considerably since the real and imaginary part of the transition amplitude $T_a+T_b$ are equal, leading to $\vv{\omega}_{1}=0$. Thus we can say that Nature chooses between two opposite directions  $\pm\vv{\omega}$, albeit between two different handed coordinate systems or chirality. Of course, we could also reinterpret equation (\ref{ErgInt}) if and only if the initial spin state is pure
 stating that the spin direction is projected onto the mixed momentum state $\omega_+\Pi_{\vv{\omega}_1+\vv{\omega}_2}+ \omega_-\Pi_{\vv{\omega}_1-\vv{\omega}_2}$. This interpretation does not hold if there is an initial correlation between spin and momentum degrees of freedom, e.g. for decay cascades (Example III).

\textbf{Example I: A spin-$\frac{1}{2}$ particle propagating through an interferometer:}
Let us consider a spin-$\frac{1}{2}$ particle that passes a beam-splitter (BS). The interaction can be reasonably well described by the unitary operator $U_{BS}=e^{-i\frac{\pi}{4}\sigma_y}$. Then the particle picks up usually a phase shift $\chi$ ($U_{phase}(\chi)=e^{-i\frac{\chi}{2}\sigma_x}$) and passes again a beam-splitter. A pure initial spin state $\rho_{spin}=\frac{1}{2}(\mathbbm{1}+\vv{s}(\theta,\phi)\cdot\vv{\sigma})$ is then changed  by a total unitary operation $U_{\textrm{IF}}=U_{BS}\cdot U_{phase}(\chi)\cdot U_{BS}$ into a final state
\begin{eqnarray}
\rho_f&=&U_{\textrm{IF}}\;\rho_{spin}\;U_{\textrm{IF}}^\dagger\nonumber\\
&=&\frac{1}{2}\{\mathbbm{1}+\left(\begin{array}{c}-s_x(\theta,\phi) \cos\chi+s_y(\theta,\phi) \sin\chi\\
s_x(\theta,\phi) \sin\chi+s_y(\theta,\phi) \cos\chi\\-s_z(\theta,\phi)\end{array}\right)\vv\sigma\}\;.\nonumber
\end{eqnarray}
A measurement in $x$- or $y$-direction reveals the interference \begin{eqnarray}I(\theta,\phi)&=&Tr(\frac{1\pm\sigma_x}{2}\rho_f)=1/2(1\mp\sin\theta\cdot\cos(\phi+\chi))\;,\nonumber\\\end{eqnarray} and therefore the visibility is in this case $\mathcal{V}=\sqrt{s_x^2+s_y^2}=\sin\theta$.  Whereas a measurement in $z$-direction reveals the probability to propagate via the upper or lower path of the interferometer\begin{eqnarray}I(\theta,\phi)&=&Tr(\frac{1\pm\sigma_z}{2}\rho_f)=1/2(1\mp\cos\theta)\;,\end{eqnarray} and therefore the ``\textit{which way}'' information is quantified by the predictability $\mathcal{P}=|-s_z|=\cos\theta$, i.e. the initial spin contribution in $z$-direction. Thus $\mathcal{V},\mathcal{P}$ can be chosen via the preparation of the initial spin (given the interferometric setup).

In the following we go one step further and assume that one has an interferometric device including a measurement along $x$- or $y$-direction, albeit revealing the ``\textit{wave particle}'' information of the spin, however, we now change the probability of the beam-splitter to a $\|T_a\|^2:\|T_b\|^2=|S|^2:|P|^2$ one and find (see Eq.~(\ref{generalresult}))
  \beq
  I(\theta,\phi)&=&(|S|^2+|P|^2)\cdot\nonumber\\
  \lefteqn{\left(1\mp \frac{2 |S| |P|}{|S|^2+|P|^2}\;\vv{n}(0,\chi_{SP})\cdot\vv{s}(\theta,\phi)\right)}
  \eeq
by choosing the following $T$ matrices
\beq
T_a=\|T_a\|\, \mathbbm{1}\quad\textrm{and}
\; T_b=\|T_b\|\, U_{\textrm{IF}}^\dagger\; \vv{n}(0,\chi_{SP})\cdot\vv{\sigma}\; U_{\textrm{IF}}\;.\nonumber
\eeq
With Eq.~(\ref{predictabilityvisibility}) we obtain a quantified quantum information theoretic interpretation of Bohr's complementarity relation of the interferometer in terms of $\mathcal{P}^2+\mathcal{V}^2=1$, i.e. how much interference contrast corresponds to a certain interferometric choice. Note the conceptual difference to the previous scenario where $\mathcal{V}$ referred to the chosen initial spin. We are therefore not limited to spin-$\frac{1}{2}$ decays since we are interested in the two amplitude process of the weak interaction. Let us apply now this view to hyperon decays.

\textbf{Example II: Spin-$\frac{1}{2}$ hyperon nonleptonic decays:}
The conservation of the total spin implies that the final states can have two different angular momentum eigenstates. We denote with $S/P$ the amplitude of the parity violating/conservating process  (corresponding to the $S,P$ wave with angular momentum $l=0,1$), then the decay matrices are $T_a=S\; \mathbbm{1}$ and $T_b=P\;\vv{n}\cdot\vv{\sigma}$ computed via the corresponding spherical harmonics and Clebsch-Gordan coefficients~\cite{ErikThesis}. The angular distribution of the normalized momentum $\vv{n}$ of the daughter particle is given by Eq.(\ref{ErgInt}) with $\vv{\omega}_1=0, \vv{\omega}_2=\vv{n}$ and $\omega_+-\omega_-=\alpha$.
Depending on the production process symmetries on $\vv{s},\vv{n}$ may be superimposed.
In  the  standard  phenomenology  of  hyperon decays
the following decay parameters are introduced~\cite{BookPilkuhn}
\begin{eqnarray}
\alpha &=&\frac{2 Re\{S^*P\}}{|S|^2+|P|^2}\;,\;
\beta\,=\,\frac{2 Im\{S^*P\}}{|S|^2+|P|^2}\,=\,\sqrt{1-\alpha^2} \sin\phi\;,\nonumber\\
\gamma &=& \frac{|S|^2-|P|^2}{|S|^2+|P|^2}\,=\,\sqrt{1-\alpha^2} \cos\phi\;.\nonumber
\end{eqnarray}
and the measured values of the parameters $\alpha,\phi$ are given in Ref.~\cite{ParticleDataBook}. We can connect these parameters to our information theoretic quantities $\mathcal{V}=\sqrt{\alpha^2+\beta^2},\mathcal{P}=|\gamma|$ and $\alpha=\mathcal{V}\cos\chi_{SP}$. Time reversal symmetry requires, in the absence of final-state interactions, that the $S$- and $P$-wave amplitudes should be relatively real, hence $\phi\approx 0$, which means that the asymmetry parameter $\alpha$ approximately equals the visibility $\mathcal{V}$ of the $S$ and $P$ wave.
\begin{center}
\begin{table*}
\renewcommand{\arraystretch}{1.1}
\begin{tabular}{||c||c|c|c|c||}
 \hhline{|t:=====:t|}
hyperon (quarks) & decay channel (frequency)& phase shift $\chi_{SP}$ in $[\pi]$& visibility $\mathcal{V}$ & predictability $\mathcal{P}$\\
\hhline{|:=====:|} $\Lambda(uds)$&$p \pi^-$ (63.9\%)& $-(0.043\pm0.023)$&$0.648\pm0.014$ &$0.762\pm0.012$\\
\hline &$n \pi^0$ (35.8\%)& $-(0.042\pm0.023)$&$0.656\pm0.040^*$ &$0.755\pm0.034^*$\\
\hline $\bar\Lambda(\bar u \bar d \bar s)$&$\bar p \pi^+$ (63.9\%)&$0.036\pm0.021$& $0.714\pm0.079^*$ &$0.700\pm0.080^*$\\
\hline $\Sigma^-(dds)$&$n \pi^-$ (99.8\%)&$-(0.38\pm0.16)$& $0.19\pm0.24$ &$0.98\pm0.05$\\
\hline $\Sigma^+(uus)$&$p \pi^0$ (51.6\%)&$-(0.038\pm0.035)$& $0.976\pm0.016$ &$0.161\pm0.097$\\
\hline &$n \pi^+$ (48.3\%)&$0.41\pm0.13$& $0.24\pm0.33$ &$0.972\pm0.078$\\
\hline $\Xi^0(uss)$&$\Lambda \pi^0$ (99.5\%)&$0.214\pm0.085$& $0.53\pm0.11$ &$0.85\pm0.07$\\
\hline $\Xi^-(dss)$&$\Lambda \pi^-$ (99.8\%)&$0.0226\pm0.0086$& $0.459\pm0.012$ &$0.8884\pm0.0062$\\
 \hhline{|b:=====:b|}
\end{tabular}
\caption{The information theoretic content of the weak decay processes violating the strangeness number. The values and errors are taken from the particle data book~\cite{ParticleDataBook}. The asterisk $^*$ denotes that the phase $\phi$ was not independently measured but deduced via conservation laws from the other decay modes. $\bar\Lambda$ also decays into $n \pi^0$, however, it has not yet been measured. The only non-resonant hyperons with spin-$\frac{3}{2}$ are the $\Omega^\pm$ particles, the measured asymmetry are close to zero ($\alpha_\Omega=0.0180\pm0.0024$) and no $\phi$ information is available.}\label{table1}
\end{table*}
\end{center}In the table~\ref{table1} we list all hyperon decays for which $\alpha$ and the phase $\phi$ have been measured. Let us remark here that if $\phi$ is not measured we cannot say anything about the visibility since it depends strongly on $\phi$ except for $\alpha$ close to $1$.  We observe that the phase shifts $\chi_{SP}$, which reduce the total visibility of the decay, are rather small. For some decays, where the phase shifts are known with limited accuracy, a larger values of the phase shifts (up to $\approx 0.5\pi$) are still possible.  The predictability of all weak decay processes is rather large except for the $\Sigma^+$ particle. This particle has --in contrary to the other hyperons-- two decay channels with nearly equal branching ratios. Due to the rule for the hyper charge $Y$ ($\Delta Y=\frac{1}{2}$) and the spurious-kaon rule~\cite{BookPilkuhn} identical $\alpha$'s and  $\phi$'s are expected for the following two pairs of decays: $\Lambda\longrightarrow \pi^-p;\pi^0 n$ and  $\Xi^{-,0}\longrightarrow \Lambda \pi^{-,0}$. That is significantly different to the case of the $\Sigma^\pm$ decays where the visibility is close to zero. This implies that one of these decays must be mainly a $S$-wave and the other one mainly a $P$-wave. This means also that the imperfection of the spin measurement is low in strong contrast to $\Xi^-$ decays where the probability of both directions is close to $\frac{1}{2}$.

In summary, the weak decay process of hyperons with spin can be viewed as an interferometer device with a fixed phase and a beam splitter with a fixed splitting. This corresponds to fixed visibilities and predictabilities independently of the initial spin of the decaying hyperon.

\textbf{Example III: Cascade of hyperon decays:}
Considering subsequent decays of hyperons into hyperons and finally into non-strange particles we can apply the transition matrices $T_\mu, T_\nu$ in a straightforward way by
\beq
I(\theta_\nu,\phi_\nu;\theta_\mu,\phi_\mu)&=& Tr\left(T_{\nu}\cdot T_{\mu}\;\rho_{spin}^\mu\;T_{\mu}^\dagger\cdot T_{\nu}^\dagger\right)\nonumber\\
&\equiv& Tr\left(K_+\rho_{spin}^\mu K_+^\dagger\right)+Tr\left(K_-\rho_{spin}^\mu K_-^\dagger\right)\;.\nonumber
\eeq
The last equation holds since we have again a complete positive evolution. Let us remark here that these Kraus operators $K_\pm$ are not a product of the Kraus operators of each decay. The reason is that after the first decay we have correlations between the spin and momentum degrees of freedom, i.e. the crucial initial condition $\rho_{mom}\otimes\rho_{spin}$ does not hold. Contrary to the dynamics of closed systems open quantum systems have no continuity of time and therefore it is not always possible to formulate the general dynamics by means of differential equations generating contractive families. On the contrary cascades allow to reveal also more information on the $S$-wave and $P$-wave coefficients, i.e. after a straightforward computation we find for the decay
of a  hyperon $\mu$ (spin-$\frac{1}{2}$) into a hyperon $\nu$ (spin-$\frac{1}{2}$) and finally into a baryon (spin-$\frac{1}{2}$) the following Kraus coefficients $\omega_\pm=\frac{1}{2}(1\pm\frac{|\vv{\tau}|}{\tau_0})$
and $\vv{\omega}_1=0,\vv{\omega}_2=\vv{\tau}$ with \beq
\tau_0&=&(|S_\mu|^2+|P_\mu|^2)(|S_\nu|^2+|P_\nu|^2)(1+\alpha_\mu \alpha_\nu (\vv{n}_\mu\cdot\vv{n}_\nu))\nonumber\\
\vv{\tau}&=&(|S_\mu|^2+|P_\mu|^2)(|S_\nu|^2+|P_\nu|^2)\nonumber\\
&&\left((\alpha_\mu +\alpha_\nu (1-\mathcal{P}_\mu)\; \vv{n}_\mu\cdot\vv{n}_\nu)\;\vv{n}_\mu\right.\nonumber\\
&&\left.\qquad+\alpha_\nu\; \mathcal{P}_\mu\; \vv{n}_\nu+\alpha_\nu\beta_\mu\; \vv{n}_\mu\times\vv{n}_\nu\right)\;.\nonumber
\eeq
The angle between the momenta of the two hyperons $\vv{n}_\mu\cdot\vv{n}_\nu$ is fixed by the request to orientate the two reference systems with respect to the initial hyperon spin. We obtain a conceptually simple interpretation of the decay cascade as in Example II: the two quantization directions are $\pm\vv{\tau}$ and the momentum direction of the baryon is measured with probability $\frac{1}{2}(1\pm\frac{|\vv{\tau}|}{\tau_0})$ along them. Notable that $\vv{\tau}$ has a contribution $(1-\mathcal{P}_\mu)$ along the direction of the momentum of the daughter hyperon $\vv{n}_\mu$, whereas the direction of the momentum of the baryon $\vv{n}_\nu$ is multiplied by the predictability $\mathcal{P}_\mu$. For example the predictabilities of the two decay cascades $\Xi^{-,0}\longrightarrow \Lambda \pi^{-,0}\longrightarrow p \pi^{-,0} \pi^-, n \pi^{-,0} \pi^0$  are large $\mathcal{P}_\mu\approx 0.8$ and therefore the quantization axis becomes to a good approximation $\approx\alpha_\mu \vv{n}_\mu+\alpha_\nu \vv{n}_\nu$.

\textbf{Entanglement:} Entangled $\Lambda\bar\Lambda$ pairs can be produced for example via proton-antiproton annihilation and the full spin structure can be experimentally determined~\cite{PS185Exp1}. In this section we analyse the entanglement properties and discuss whether Bell's nonlocality or contextuality can be revealed in such experiments. Let us assume that (i) there is no initial correlation between the momentum degrees of freedom and the spin degrees of freedom and (ii) there is no entanglement between the momentum degrees of freedom. Our formalism extends straightforwardly to the two-particle case by taking the tensor product of the Kraus operators.  Experiments~\cite{PS185Exp1,PS185Exp2,PS185Exp3} suggest that the initial spin state is a maximally entangled Bell state (except for backward scattering angles). Therefore without loss of generality we can choose the antisymmetric Bell state $|\psi^-\rangle= \frac{1}{\sqrt{2}}\{|\Uparrow\Downarrow\rangle-|\Downarrow\Uparrow\rangle\}$ and the angular momentum distribution becomes
\beq
I(\theta_1,\phi_1;\theta_2,\phi_2)&=& \frac{(|S|^2+|P|^2)^2}{4}\left\lbrace 1-\alpha_\Lambda \alpha_{\bar \Lambda}\; \vv{n}_1\cdot\vv{n}_2\right\rbrace\;.\nonumber\\
\eeq
Since $\vv{n}_1\cdot\vv{n}_2$ is multiplied by a constant $\alpha_\Lambda \alpha_{\bar \Lambda}$ T\"ornquist~\cite{Tornquist} concluded that $\Lambda$ decays ``\textit{as if it had a polarization $\alpha_\Lambda$ tagged in the direction of the $\pi^+$ (coming from the $\bar\Lambda$) and vice versa}''. The knowledge of how one of the $\Lambda's$ decayed --or shall decay (since time ordering is not relevant)-- reveals the polarization of the second $\Lambda$. He concludes that this is the well-known Einstein-Podolsky-Rosen scenario.

\begin{figure}
\center{(a)\includegraphics[width=0.2\textwidth,
keepaspectratio=true]{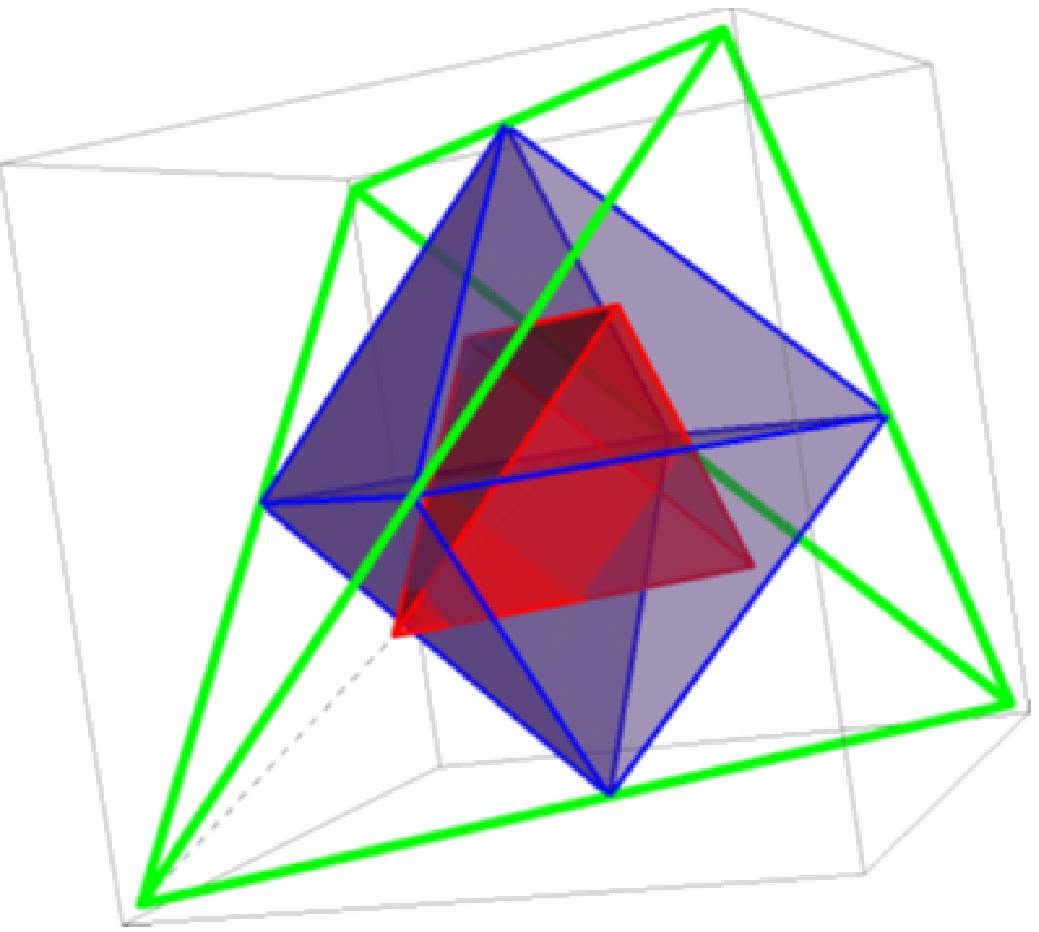}
(b)\includegraphics[width=0.22\textwidth,
keepaspectratio=true]{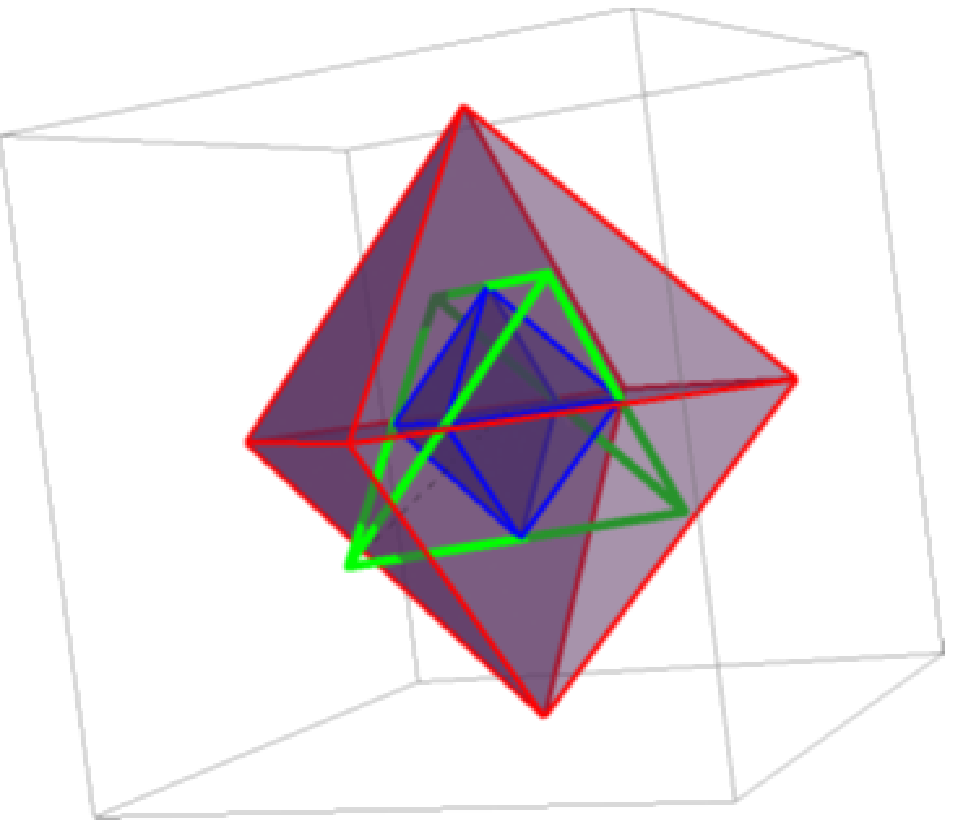}\caption{(Color online) The big (green) tetrahedron corresponds to the geometrical illustration of all local maximally mixed bipartite qubit states which can be represented by three real numbers. The four corners represent the four maximally entangled Bell states which are the only pure states in the picture. Separable states are those inside the (blue) double pyramid where the surfaces correspond to optimal entanglement witnesses. For the $\Lambda\bar\Lambda$ system we can interpret that (a) the Hilbert space shrinks with the factor $\alpha_\Lambda\alpha_{\bar\Lambda}$ visualized by the small (red) tetrahedron. Since the corners are still outside of the optimal entanglement witnesses (double pyramid), entanglement can be detected via the weak decaying process. Or equivalently, as illustrated in (b) we could also say that weak interactions correspond to imperfect spin measurements blowing up the optimal entanglement witnesses, huge (red) double pyramid.
}\label{tetrahedron}}
\end{figure}

In general entanglement is detected by a certain observable that can witness the entanglement content, i.e. a Hermitian operator $\mathcal{W}$ for which holds $Tr(\mathcal{W}\rho)<0$ for at least one state $\rho$ and $Tr(\mathcal{W}\rho_{sep})\geq 0$ for all separable states $\rho_{sep}$. For the antisymmetric Bell state such an optical entanglement witness is given by $\mathcal{W}=
\frac{1}{3}(\mathbbm{1}\otimes\mathbbm{1}+\sum_i \sigma_i\otimes\sigma_i)
$ (any other witness can be obtained by local unitary transformations). Since the weak interaction only allows for an imperfect spin measurement we have to multiply the spin part by $\alpha_\Lambda \alpha_{\bar \Lambda}$. Thus $Tr(\mathcal{W}_{\alpha}\rho_{\Lambda\bar\Lambda})$ results in
\beq\frac{1}{3}-\alpha_\Lambda \alpha_{\bar \Lambda}\geq 0\;\forall\;\rho_{sep}\;,\eeq
which is clearly violated since $\alpha_\Lambda \alpha_{\bar \Lambda}=0.46\pm0.06$~\cite{ParticleDataBook}. Therefore, the measurement of the correlation functions $\langle\sigma_i\otimes\sigma_i\rangle$ in $x,x$ and $y,y$ and $z,z$ directions of the $\Lambda$ and $\bar\Lambda$ reveals entanglement. Generally, we can say that the asymmetries lead to imperfect spin measurements which shrink the observable space. Equivalently, we can say that the given interferometric device leads to a shrinking of the Hilbert space of the accessible spin states. In Fig.~\ref{tetrahedron} we have visualized the geometry of the Hilbert space for all locally mixed states of bipartite qubits forming a magic simplex~\cite{BHN1,BHN2,BHN3}. Locally mixed states are those for which any partial trace reduces to the maximally mixed state. This set of states can be described by three real numbers. Positivity requires that these three numbers form a tetrahedron with the four maximally entangled Bell states in the corners. The separability condition corresponds to a double pyramid. The surfaces of the pyramid correspond to the optimal entanglement witnesses. As shown in Fig.~\ref{tetrahedron} the factor $\alpha_\Lambda \alpha_{\bar \Lambda}$ (a) shrinks the total state space (smaller red tetrahedron) or (b) blows up the optimal entanglement witnesses. Since $\alpha_\Lambda \alpha_{\bar \Lambda}>\frac{1}{3}$ one can distinguish between entangled and separable states directly, i.e. without the additional information coming from the two interferometer devices.

\textbf{Testing for local realism or contextuality:} The next question is whether the entanglement observed in this system manifests itself in the most counterintuitive properties of quantum theory, i.e. Bell's nonlocality or contextuality. The notion of contextuality, introduced by John Bell~\cite{BellContextuality}  and by Kochen and Specker~\cite{KochenSpecker}, can be explained as follows. Suppose that a measurement $A$ can be jointly performed with either measurement $B$ or $C$, i.e. without disturbing the measurement $A$. Measurements $B$ and $C$ are said to provide a context for the measurement $A$. The measurement $A$ is contextual if its outcome depends on whether it was performed together with $B$ or with $C$. Therefore, the essence of contextuality is the lack of possibility to assign an outcome to A prior to its measurement and independently of the context in which it was performed. The seemingly different Bell theorem is in fact a special instance of the Kochen-Specker theorem where contexts naturally arise from the spatial separation of measurements. The usual toolboxes for revealing nonlocality and contextuality are state in-dependent or dependent inequalities of probabilities. In the following we analyze these inequalities for the hyperon-antihyperon system.

We considered  Bell inequalities for qubits with two, three or four different choices of observables for both particles, i.e. the following expressions
\beq
I_{2}&=& Prob(\vv{a}_1, \vv{b}_1)+Prob(\vv{a}_1, \vv{b}_2)+Prob(\vv{a}_2, \vv{b}_1)\nonumber\\&&-Prob(\vv{a}_2, \vv{b}_2)-
Prob(\vv{a}_1)-Prob(\vv{b}_1)\;,\nonumber\\
I_{3}&=& Prob(\vv{a}_1, \vv{b}_1)+Prob(\vv{a}_1, \vv{b}_2)+Prob(\vv{a}_1, \vv{b}_3)\nonumber\\&&+Prob(\vv{a}_2, \vv{b}_1)+Prob(\vv{a}_2, \vv{b}_2)-Prob(\vv{a}_2, \vv{b}_3)\nonumber\\
&&+Prob(\vv{a}_3, \vv{b}_1)-Prob(\vv{a}_3, \vv{b}_2)-
Prob(\vv{a}_1)\nonumber\\&&-2 Prob(\vv{b}_1)- Prob(\vv{b}_2)\;,\nonumber
\eeq
\beq
I_{4}&=& Prob(\vv{a}_1, \vv{b}_1)+Prob(\vv{a}_1, \vv{b}_2)+Prob(\vv{a}_1, \vv{b}_3)\nonumber\\&&+Prob(\vv{a}_1, \vv{b}_4)+Prob(\vv{a}_2, \vv{b}_1)+Prob(\vv{a}_2, \vv{b}_2)\nonumber\\
&&+Prob(\vv{a}_2, \vv{b}_3)-Prob(\vv{a}_2, \vv{b}_4)+Prob(\vv{a}_3, \vv{b}_1)\nonumber\\&&+Prob(\vv{a}_3, \vv{b}_2)-Prob(\vv{a}_3, \vv{b}_3)+Prob(\vv{a}_4, \vv{b}_1)\nonumber\\
&&-Prob(\vv{a}_4, \vv{b}_2)-
Prob(\vv{a}_1)-3 Prob(\vv{b}_1)\nonumber\\&&- 2Prob(\vv{b}_2)- Prob(\vv{b}_3)\;,\nonumber
\eeq
where $Prob(\vv{a}_i, \vv{b}_j)$ is the probability that the first particle is measured along  $\vv{a}_i$ giving e.g. the plus result and the second particle is measured along the direction $\protect\vv{b}_j$  and gives e.g. the plus result. For any local realistic theories $I_i\leq 0$ has to hold. We find that the strongest constraint is found for the famous CHSH-Bell inequality~\cite{chsh} ($I_2$), i.e. \beq
\alpha_\Lambda\alpha_{\bar\Lambda}\approx (0.46\pm0.06)\leq\frac{1}{\sqrt{2}}\eeq  that has to hold for all local realistic theories. Clearly,  this reveals no violation for the $\Lambda\bar\Lambda$ system. However, we have to remark that for testing theories based on local hidden parameters versus the predictions of quantum mechanics there are two important requirements: Firstly, one is not allowed to refer to results that are deduced from quantum mechanical considerations. In our case this means that we cannot re-normalize to the asymmetry parameters $\alpha_{\Lambda,\bar \Lambda}$ since they are obtained from quantum mechanical properties. Secondly, any conclusive Bell's test requires that an active control of the experimenter about measurement setting is given, i.e. a choice of freedom which property of the state will be measured. Otherwise it is straightforward to construct a local realistic hidden variable theory to explain the data. This active control is clearly not available for weakly decaying hyperons since the quantization axes are spontaneously chosen. Some authors~\cite{Chen:2013} argued that one can circumvent the requirement since an observer can choose a coordinate system at will at each side. We disagree since one needs always one common reference system to describe the two-particle decay. Note that if there would be entanglement in the momentum state then this entanglement can (in some cases) be transferred to the spin states for a relativistically boosted observer~\cite{RelEnt}. However, in this case also the operators corresponding to the measurements are boosted in exactly such a way that the expectation value is independent of the considered reference frame in agreement with special relativity. Hence a violation of a Bell inequality found in one reference system is also violated in another one. In summary, in contrary to the weakly decaying spinless K-meson system~\cite{Hiesmayr:2012} the active measurement procedure for hyperons decays are not available and therefore no contradiction to the premisses of local realism based on Bell inequalities can be derived.

Contextuality tests usually do not suffer from these requirements.
Due to the imperfect spin measurements we expect for the hyperons a decrease of probabilities and therefore a dependence of the violation on the asymmetry term. The tests need the property of joint measurements that is available for the hyperon system since we have the tensor product between the observables. If one follows the interpretation that the accessible state space is shrunken (that holds true only if there is no initial correlation between the momentum degrees of freedom and the spin degrees of freedom), then all state independent proofs such as the Mermin-Peres square~\cite{Mermin,Peres,contextualityreview} hold also for the $\Lambda\bar\Lambda$ system and non-contextuality is revealed. Applying the other interpretation that holds also in the general case (see example III) the observables have to be multiplied by $\alpha_\Lambda$ or $\alpha_{\bar\Lambda}$, respectively, and the Mermin-Peres inequality leads to
\beq
I_{\textrm{contextuality}}=(\alpha_\Lambda^2+\alpha_{\bar\Lambda}^2)^2+2\,\alpha_\Lambda^3\alpha_{\bar\Lambda}^3\leq 4\;.
\eeq
which is not violated. Assuming $\mathcal{CP}$ conservation contextuality would be revealed if $\alpha>0.88$ and hence greater than the bound from the Bell inequality $\alpha>0.84$.

\textbf{Conclusion and Summary:} In this contribution we introduce an information theoretic approach to hyperons decaying via the weak interaction. The parity violating and non-violating processes can be considered as two different paths in an interferometer. We find that weak interaction chooses for each hyperon an interferometric device with a fixed visibility, i.e. interference contrast, and a fixed phase shift. The visibility results from a non-symmetric beam-splitter. Based on Bohr's complementarity relation we can derive the related predictability, quantifying the ``\textit{particle-like}'' property, which turns out to be in general high (except for one decay mode of the $\Sigma^+$ hyperon). Thus weak interaction distinguishes strongly between processes conserving and not conserving the parity symmetry.

Applying the open quantum formalism we find a simple and transparent method to describe weakly decaying hyperons. We find that the decay via the weak interaction corresponds to an imperfect spin measurement, where right and left-handed coordinate systems mix. 

Equipped with this quantum information theoretic knowledge we proceeded to two-particle system, e.g. the $\Lambda\bar\Lambda$ system. Entanglement can be proven experimentally although the imperfection in the spin measurements. We show further that it is impossible to find a conclusive contradiction between quantum theory and local realistic theories via Bell inequalities since weak decays offer no active control over the quantization directions. Last but not least we investigate contextuality in the $\Lambda\bar\Lambda$ system. We find that it cannot be revealed by the Mermin-Peres square if one assumes that the violation of the parity symmetry shrinks the accessible observable space. The second possible interpretation ---violation of the parity symmetry corresponds to the shrinking of accessible state space--- would reveal the contextual nature of quantum theory, however, this interpretation does not hold if there are correlations between the spin and momentum degrees of freedom. In opposition two the weakly decaying K-meson system the observation of the violation of the $\mathcal{CP}$ symmetry would not considerably alter the results.

We believe that the presented information theoretic analyzes of hyperon decays and entanglement helps to bring the data from the upcoming experiments in a unified picture and contributes to the understanding of weak interactions.

\textbf{Acknowledgements:}  The author is grateful to Andzej Kupsc (Uppsala University) for introducing her to the $\Lambda\bar\Lambda$ system and many fruitful discussions and comments to the paper. She also thanks Reinhold A. Bertlmann (University of Vienna) for discussions and gratefully acknowledges the Austrian Science Fund (FWF-P26783).

\end{document}